\documentclass[aps,10pt,prx,twocolumn,superscriptaddress,noeprint,
longbibliography,floatfix]{revtex4-2}

\usepackage{graphicx}
\usepackage{bm}
\usepackage{amsmath, amssymb}
\usepackage{booktabs}
\usepackage{mathrsfs}
\usepackage{makecell}
\usepackage[normalem]{ulem}
\usepackage{pifont}
\usepackage{placeins}
\usepackage{lipsum}
\usepackage{wasysym}

\usepackage[mathscr]{euscript}
\usepackage{multirow}
\usepackage{physics}
\usepackage{xcolor}
\usepackage{siunitx}
\usepackage{placeins}
\usepackage{blindtext}
\usepackage{titlesec}
\usepackage{nicefrac}
\usepackage{appendix}
\definecolor{darkblue}{rgb}{0,0,.65}
\definecolor{darkgreen}{rgb}{0.28,0.41,0.19}
\definecolor{nicegreen}{rgb}{0.28,0.85,0.19}
\usepackage{mathtools}

\usepackage[
  pdfauthor={Robin Schaefer},
  pdfstartview=FitH,
  breaklinks=true,
  bookmarks=true,
  colorlinks=true,  
  linkcolor=darkblue,
  citecolor=darkblue,
  urlcolor=darkblue,
  anchorcolor=black,
  filecolor=black,
  menucolor=black
]{hyperref}
\usepackage{orcidlink}

\newcommand{\eg}{\textit{e.g.}}
\newcommand{\I}{\mathrm{i}}
\newcommand{\TMC}{T_{\rm MC} }

\newcommand{\figref}[2]{\hyperref[#1]{\autoref*{#1}(#2)}}
\newcommand{\aref}[1]{\hyperref[#1]{App.~\ref*{#1}}}

\newcommand{\kb}{k_{\mathrm{B}}}

\renewcommand{\vec}[1]{\mathbf{ #1}}

\def\equationautorefname~#1\null{Eq. (#1)\null}

\newcommand{\appref}[1]{\hyperref[#1]{App.~\ref*{#1}}}
\renewcommand{\eg}[1]{\textit{e.g.}}
\usepackage[all]{hypcap} 
\usepackage{physics}
\renewcommand{\ket}[1]{\vert #1\rangle}

\renewcommand{\expval}[1]{\langle\,#1\,\rangle}

\newcommand{\canted}{c-120$^{\circ}$}

\begin{document}

\title{Thermodynamics of the Heisenberg antiferromagnet on the maple-leaf lattice}
\author{Robin Sch\"afer\,\orcidlink{0000-0001-9728-2371}}
\email{robin.schaefer@helmholtz-berlin.de}
\affiliation{Helmholtz-Zentrum Berlin f\"ur Materialien und Energie, Hahn-Meitner-Platz 1, 14109 Berlin, Germany}
\affiliation{Dahlem Center for Complex Quantum Systems and Fachbereich Physik, Freie Universit\"at Berlin, Arnimallee 14, 14195 Berlin, Germany}

\author{Paul L. Ebert\,\orcidlink{0000-0003-1614-6920}}
\affiliation{Max Planck Institute for the Physics of Complex Systems, Noethnitzer Str. 38, 01187 Dresden, Germany}

\author{Noah Hassan}
\affiliation{Helmholtz-Zentrum Berlin f\"ur Materialien und Energie, Hahn-Meitner-Platz 1, 14109 Berlin, Germany}
\affiliation{Dahlem Center for Complex Quantum Systems and Fachbereich Physik, Freie Universit\"at Berlin, Arnimallee 14, 14195 Berlin, Germany}

\author{Johannes Reuther}
\affiliation{Helmholtz-Zentrum Berlin f\"ur Materialien und Energie, Hahn-Meitner-Platz 1, 14109 Berlin, Germany}
\affiliation{Dahlem Center for Complex Quantum Systems and Fachbereich Physik, Freie Universit\"at Berlin, Arnimallee 14, 14195 Berlin, Germany}

\author{David J. Luitz \orcidlink{0000-0003-0099-5696}}
\affiliation{Institute of Physics, University of Bonn, Nussallee 12, 53115 Bonn, Germany}

\author{Alexander Wietek}
\affiliation{Max Planck Institute for the Physics of Complex Systems, Noethnitzer Str. 38, 01187 Dresden, Germany}
\date{\today}

\begin{abstract}
We study the Heisenberg antiferromagnet on the maple-leaf lattice using several numerical approaches, focusing on the numerical linked-cluster expansion (NLCE), which exhibits an unconventional convergence extending to low and even zero temperatures. We evaluate thermodynamic properties as well as spin-spin correlations through the equal-time structure factor. Within NLCE the specific heat capacity reveals a two-peak structure at $T_1 \approx 0.479\,J$ and $T_2 \approx 0.131\,J$, reminiscent of the corresponding result for the triangular lattice. At intermediate temperatures, the spin-spin structure factor develops features that reflect the absence of reflection symmetry in the lattice. The zero-temperature convergence of NLCE enables reliable estimates of the ground-state energy and points to a short-range correlated paramagnetic ground state composed of resonating hexagonal motifs. The NLCE results are benchmarked against Pseudo-Majorana Functional Renormalization Group, finite-temperature Lanczos, and classical Monte Carlo simulations.
\end{abstract}
\maketitle

\section{Introduction}
Frustrated lattice models host exotic states of matter, such as quantum spin liquids, which evade Landau’s traditional classification of phases~\cite{savary_quantum_2017}. These systems support exotic, fractionalized quasiparticles, making them highly attractive for both theoretical and experimental investigations~\cite{smith_experimental_2025}. Frustration can arise either from anisotropic exchange couplings, as in Kitaev’s honeycomb model~\cite{kitaev_anyons_2006}, or from geometric constraints in antiferromagnetic lattices. Prominent examples of the latter include the two-dimensional kagome lattice~\cite{norman_colloquium_2016} and the three-dimensional pyrochlore lattice~\cite{hermle_pyrochlore_2004}.

\begin{figure}[t]
        \includegraphics[width=\linewidth]{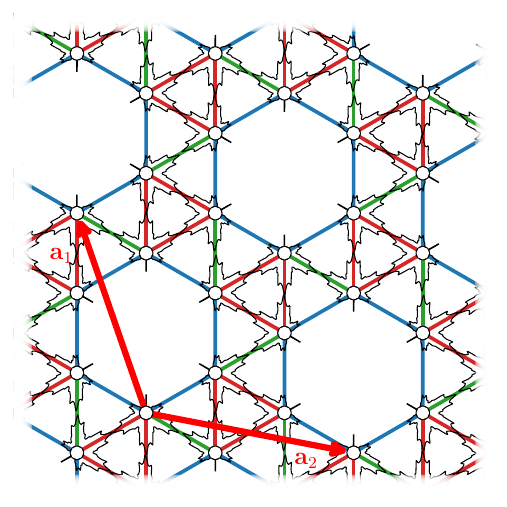}
        \caption{Depiction of the maple-leaf lattice with its three symmetry-inequivalent nearest-neighbor bonds: triangular (red), hexagonal (blue), and diagonal (green). The figure also shows the Bravais lattice vectors $\vec{a}_1$ and $\vec{a}_2$. The lattice derives its name from its coordination number of five, which allows the use of maple leaves (shown in black) to highlight the nearest-neighbor structure.}
        \label{fig:maple_leaf}
\end{figure}
More recently, another frustrated two-dimensional lattice has attracted increasing attention: the so-called maple-leaf lattice shown in \autoref{fig:maple_leaf}~\cite{betts:1995}. It has emerged as a promising platform for realizing spin-liquid behavior~\cite{sonnenschein_candidate_2024,ghosh_where_2024} or a deconfined quantum critical point~\cite{ghosh_where_2024}.
The lattice hosts three inequivalent bonds shown by different colors in \autoref{fig:maple_leaf}, leading to a variety of competing ground-state candidates at zero temperature.
For dominant diagonal bonds (green), the model exhibits a well-established dimerized phase~\cite{ghosh:2022, farnell:2011, beck:2024, gresista:2023, schmoll_bathing_2025,nyckees_tensor_2025}, reminiscent of the Shastry–Sutherland model~\cite{koga:2000} in the sense that the dimer-product state is an exact eigenstate. The ``plaquette phase'' of the Shastry–Sutherland model, breaking the $\mathbb{Z}_2$ symmetry, would correspond to a resonating hexagon state (without symmetry breaking) which is recovered for dominant hexagonal couplings (blue).
Away from the extreme limits, the nature of the ground state is still an open question. A wide range of methods, including coupled-cluster theory~\cite{farnell:2011, richter:2004, farnell:2018, farnell:2014}, exact diagonalization (ED)~\cite{farnell:2011, schulenburg:2000, schmalfuss:2002, richter:2004}, linear spin-wave theory~\cite{schmalfuss:2002}, plaquette-triplon mean-field analyses~\cite{ghosh:2024b}, pseudo-fermion functional renormalization group~\cite{gresista:2023,gresista_unconventional_2025}, neural quantum states~\cite{beck:2024}, density matrix renormalization group~\cite{ghosh_where_2024,ghosh:2024c}, infinite density matrix renormalization group (iDMRG)~\cite{beck:2024, gembe:2024}, bond-operator mean-field theory~\cite{ghosh_where_2024}, and infinite projected entangled pair states (iPEPS) calculations~\cite{schmoll_bathing_2025,nyckees_tensor_2025} outline various possibilities for the ground state.

In particular, the isotropic point remains under debate, with proposals ranging from a dressed version of the ``canted 120$^{\circ}$’’ (\canted{}) order realized by the classical model~\cite{farnell:2011, gembe:2024, ghosh:2025,nyckees_tensor_2025} or a paramagnetic ground state~\cite{schmoll_bathing_2025}. Ref.~\cite{sonnenschein_candidate_2024} discusses the possibilities of different quantum liquids. Ongoing ED study suggests the existence of the magnetically ordered state~\cite{ebert:2026}. Only a few works focus on finite-temperature properties~\cite{ghosh:2025, hutak_thermodynamics_2025}.

On the experimental front, several synthetic compounds~\cite{cave:2006, aliev:2012, venkatesh:2020, saha:2023, Aguilar-Maldonado:2025} as well as natural minerals~\cite{haraguchi:2021, haraguchi:2018, mills:2014, fennell:2011} have been found to approximately realize the maple-leaf lattice. Experiments have focused on magnetization plateaus~\cite{schmoll:2024a, schmoll_bathing_2025, ghosh:2023, farnell:2011, richter:2004}, the origin of Bonner–Fisher–like behavior~\cite{haraguchi:2021, makuta:2021, ghosh:2024c, venkatesh:2020}, and excitations above the dimer state~\cite{ghosh:2023, esaki:2025}.

In light of the growing experimental opportunities and the lack of theoretical understanding, we employ a variety of numerical methods to investigate the thermodynamic properties of the isotropic Heisenberg antiferromagnet. Specifically, we utilize the numerical linked-cluster expansion (NLCE) algorithm~\cite{rigol_numerical_2006,rigol_numerical_2007,tang_a_2013,schaefer_pyrochlore_2020,schaefer_magnetic_2022}, pseudo-Majorana functional renormalization group (PMFRG) techniques \cite{niggemann2021frustrated, schneider2024temperature}, the finite-temperature Lanczos method (FTLM)~\cite{jaklic:1994,sugiura:2012, sugiura:2013,schnack_magnetism_2018, wietek:2019,schaefer_pyrochlore_2020}, and classical Monte Carlo simulations~\cite{metropolis_equation_1953}. In addition to finite-temperature behavior, our approach provides insights into the zero-temperature phase. 

Our main findings are as follows:
(i) Our NLCE results exhibit a double-peak structure, indicating the emergence of two dominant temperature scales.
(ii) We obtain a reliable estimate of the equal-time structure factor at intermediate temperature in the thermodynamic limit.
(iii) The finite-temperature calculations allow an estimate of the ground-state energy.
(iv) The convergence of the NLCE suggests a simple paramagnetic ground state based on weakly coupled hexagons.

The outline of this paper is as follows: In \autoref{sec:model_and_methds}, we introduce the lattice and the numerical methods. In \autoref{sec:thermo}, we discuss the thermodynamic properties, focusing on the specific heat capacity, entropy, and energy. Finally, the structure factor and real-space correlations are analyzed in \autoref{sec:StructureFactor}.

\section{Model and methods}\label{sec:model_and_methds}The maple lattice is obtained by a 1/7 depletion of sites from the triangular lattice. It consists of a six-site unit cell arranged on a triangular Bravais lattice with a coordination number of five. There are three inequivalent bond types, which we refer to as hexagonal, triangular, and diagonal bonds, shown in blue, red, and green in \autoref{fig:maple_leaf}. The lattice can be constructed from the respective geometries: a single hexagon, two triangles, and three diagonal dimers. The lattice exhibits sixfold rotational ($C_6$) symmetry but lacks reflection symmetry as opposed to the triangular and kagome lattices. The lack of reflection symmetry is a distinctive feature of the maple leaf lattice, with implications to possible emergent orders. Following the notation of Ref.~\cite{sonnenschein_candidate_2024}, the primitive lattice vectors are given by
$\vec{a}_1=\left(\sqrt{\frac{27}{4}},-\frac{1}{2}\right)$ and $\vec{a}_2=\left(-\sqrt{\frac{3}{4}},\frac{5}{2}\right)$.
The $K$ points of the underlying triangular lattice are located at $\left(0,\pm \frac{4\pi}{3}\right)$ and $\left(\pm \frac{2\pi}{\sqrt{3}},\pm\frac{2\pi}{3}\right)$.

We focus on the isotropic nearest-neighbor Heisenberg antiferromagnet with $J>0$:
\begin{align}
    H = J \sum_{\langle i,j \rangle} \Vec{S}_i\cdot \Vec{S}_j\,.\label{eq:H}
\end{align}

We discuss the individual methods used in this work in the remainder of this section. Readers interested only in their applications may skip directly to the following section.
\paragraph*{(i) Numerical linked cluster expansion.}
The NLCE algorithm is a finite temperature expansion method that enables the computation of thermodynamic observables directly in the thermodynamic limit~\cite{rigol_numerical_2006,rigol_numerical_2007}. It is based on the full diagonalization of all finite subclusters of the lattice up to a given size controlling the convergence at low temperatures.
The method is not affected by frustration or dimensionality, making it an ideal tool to study two- and three-dimensional frustrated magnets. Its broad applicability also makes it a powerful method for supporting experiments~\cite{smith_case_2022,yahne_dipolar_2024,smith_two_2025}. For an introduction, see Refs.~\cite{tang_a_2013,schaefer_pyrochlore_2020,schaefer_magnetic_2022}.

Starting from an initial building block (or a set of building blocks), the algorithm systematically constructs larger clusters by “gluing” these blocks together, thereby generating progressively larger subclusters of the lattice. These larger subclusters capture longer-range correlations, enabling calculations to reach lower temperatures. \

The expansion order of the algorithm refers to the number of building blocks taken into account. At the $n$th order, the algorithm generates subclusters $c\in\mathcal{T}_n$ that contain $n$ blocks. $\mathcal{T}_n$ refers to a collection of topologically equivalent clusters containing $n$ building blocks. Each cluster $c$ has a multiplicity $l_c$, which is an integer counting the number of distinct embeddings of the cluster in the lattice modulo translations.

For a given observable $O$, e.g., the specific heat capacity at temperature $T$, each cluster is assigned a weight $W_c$. This weight is defined as the observable evaluated on the cluster, $O_c$, minus the contributions of all smaller subclusters $s$ that can be embedded within $c$ (denoted as $s \in c$). These subclusters correspond to clusters generated in lower orders:
\begin{align}
W_c = O_c - \sum_{s \in c} W_s \,.\label{eq:weight}
\end{align}

In this way, including the cluster $c$ in the expansion ensures that only contributions not already accounted for at lower orders are added. The observable at a given order, $O\big\vert_n$, is then obtained by summing over the multiplicities and weights of all clusters:
\begin{align}
    O\big\vert_n = \sum_{k=1}^n\sum_{c\in\mathcal{T}_k} l_c W_c \,.\label{eq:NLCE}
\end{align}

A common technique to improve convergence in NLCE is the use of resummation algorithms, which take the bare NLCE result $O\big\vert_n$ and perform a resummation to unlock lower temperatures. Two widely used resummation methods are Euler’s transformation and Wynn’s algorithm~\cite{hamming_numerical_1987,william_numerical_1992,weniger_numerical_2003}. In the following, we employ Euler’s algorithm.

The algorithm also allows the computation of real-space correlation functions, which can be used to evaluate the equal-time structure factor. For this, we consider the two-point correlation function at a given distance $\vec{r}$ and inverse temperature $\beta$: $\langle \vec{S}_i \cdot \vec{S}_j \rangle_\beta$, with $\vec{r} = \vec{r}_j - \vec{r}_i$. In this case, \autoref{eq:weight} and \autoref{eq:NLCE} are no longer directly applicable, since the observable depends explicitly on spatial coordinates. For example, instead of simply multiplying by the cluster multiplicity $l_c$, one must account for the coordinates of each embedding individually to correctly include all contributions. A detailed explanation can be found in the appendix of Ref.~\cite{smith_single_2025}.

We have implemented and tested eight different approaches using various building blocks, of which we discuss two in detail. Other expansions are briefly discussed in \autoref{app:NLCE}. The first expansion is hexagon-based, employing a single hexagonal building block translated according to the triangular Bravais lattice. Neighboring hexagons are coupled via two triangular (red) and one diagonal bond (green).

In the second expansion, we employ edge-sharing triangles as building blocks. There are eight inequivalent triangles modulo translations. In contrast to the hexagon expansion, it is necessary to include a zeroth order to properly account for edge-sharing units, as each edge directly contributes to the energy. The zeroth order consists of a single edge, which is subtracted from each cluster for every edge it contains. This correction must also be considered -- with its directional dependence -- when computing the real-space correlations. Further details can be found in \autoref{app:NLCE}.

\paragraph*{(ii) Pseudo-Majorana Functional Renormalization Group.}
PMFRG is a powerful tool to calculate spin correlations for a wide range of quantum spin systems. It is based upon a representation of spins in terms of real Majorana fermions~\cite{Martin1959,Tsvelik1992}
\begin{equation}
    S^\alpha = -\frac{\I}{2}\epsilon_{\alpha\beta\gamma}\eta^{\beta}\eta^{\gamma}\,,
\end{equation}
where the Majorana operators $\eta^\alpha$ (with $\alpha=x,y,z$) fulfill the relations $(\eta^\alpha)^\dagger=\eta^\alpha$ and $\{\eta^\alpha,\eta^\beta\}=\delta_{\alpha\beta}$. The PMFRG is an improved version of the Pseudo-Fermion Functional Renormalization Group (PFFRG), which expresses the spin operators in terms of {\it complex} Abrikosov fermions. The PFFRG has been applied to a variety of frustrated spin systems in the past years~\cite{reuther2010j, muller2024pseudo}. However, due to the appearance of unphysical states related to the Abrikosov fermion representation, non-zero temperatures could not be investigated. The PMFRG employed here does not introduce unphysical states and is therefore well suited to be applied at finite temperatures \cite{niggemann2021frustrated, niggemann2022quantitative}.

At its core, the PMFRG presents a hierarchy of coupled differential equations in the Majorana $N$-point vertex functions \cite{kopietz2010introduction} in terms of derivatives with respect to a flow parameter. 
Generally, the flow parameter can be chosen in two ways: as a physical low-energy cutoff ($\Lambda$-flow) or (upon a small redefinition of Fourier transformed fields) as temperature $T$ itself \cite{schneider2024temperature}. Since the $T$-flow method allows us to calculate spin-correlations within wide temperature ranges in a single numerical run, we apply this approach here. To solve the flow equations, one needs to truncate the hierarchy at a certain order, usually by setting the 6-point vertex and higher vertices to zero. Higher order corrections that are necessary to stabilize the flow at low values of the flow parameter can be added via the utilization of the so-called Katanin truncation scheme, which self-consistently couples the self-energy to the 4-point vertex \cite{katanin2004fulfillment, reuther2010j}.
Having obtained the 4-point vertex $\Gamma_{ij}^{xyxy}(\I \nu_n,\I\omega_n,\I\omega_n')$ the frequency dependent spin-spin correlations can be calculated via
\begin{equation}
    \begin{aligned}
        \expval{S^z(\I\nu_n)&S^z(0)}_\beta = -\sum_{\I\omega_n}G(\I \omega_n)G(\I\omega_n + \I\nu_n)\delta_{ij}\\
        + \sum_{\I\omega_n, \I\omega_n'} &\bigg[ G(\I\omega_n)G(\I\omega_n + \I\nu_n)G(\I\omega_n')G(\I\omega'_n + \I\nu_n) \, \times\\
        &\Gamma^{xyxy}_{ij}(\I\nu_n, \I\nu_n + \I \omega_n + \I\omega_n', \I\omega_n - \I\omega_n')\bigg]\,,
    \end{aligned}
\end{equation}
where $\I\omega_n,\I\omega_n'$ are fermionic and $\I \nu_n$ bosonic Matsubara frequencies. Furthermore, $G(\I\omega_n)=(\I\omega_n\sqrt{T} -\Sigma(\I\omega_n))^{-1}$ is the (dressed) Majorana Green's function, $\Sigma(\I\omega_n)$ is the self-energy (2-point vertex), and the factor $\sqrt{T}$ introduces the flow parameter in the $T$-flow scheme. 

To obtain the equal-time spin correlations, we sum over all Matsubara frequencies
\begin{equation}
    \expval{S^z(t=0)S^z(0)}_\beta = \sum_{\I\nu_n}\expval{S^z(\I \nu_n)S^z(0)}_\beta\,.\label{eq:equal_time_PMFRG}
\end{equation}

Our real-space approximation consists of limiting correlation lengths to 12 lattice spacings and setting longer-distance spin correlations to zero. Furthermore, we use a maximum of 42 discrete Matsubara frequency points to express the frequency dependence of the vertex functions.

\paragraph*{(iii) Exact diagonalization.} 
    We perform Lanczos iterations on all symmetry sectors labeled by $S^z_{\mathrm{tot}} = \sum_i S_i^z$ as well as an irreducible representation of the space group~\cite{wietek:2025:xdiag}.
    Once the Lanczos routine converges, it contains information about the whole spectrum that can be used to compute finite temperature expectation values through what is known as the \emph{finite temperature Lanczos method}~\cite{schnack_magnetism_2018, jaklic:1994} (FTLM) or \emph{thermal pure quantum} (TPQ) \emph{states}~\cite{sugiura:2012, sugiura:2013, wietek:2019, schaefer_pyrochlore_2020}.
    In short, this approach exploits two approximations at the same time: First, traces can be efficiently evaluated in a Monte-Carlo fashion using the ``random-trace estimator''~\cite{girard:1989, hutchinson:1989}
    \begin{equation}
        \Tr[A] = \dim[A] \, \mathbb{E}[\langle r\vert  A \vert  r \rangle]\,,
    \end{equation}
    where $\mathbb{E}$ denotes the expectation value, and $\vert r\rangle$ is a normalized random vector distributed such that $\mathbb{E}[\vert r\rangle\langle r\vert ] = \mathrm{id}/\dim[A]$.
    Hence, a canonical expectation value becomes 
    \begin{equation}\label{eq:tpq_estimator}
        \langle A \rangle_{\beta} = \frac{\Tr[A e^{-\beta H}]}{Z} = \frac{\mathbb{E}[\langle r_\beta \vert  A \vert  r_\beta \rangle]}{\mathbb{E}[\langle r_\beta \vert r_\beta \rangle]}\,,
    \end{equation}
    where we introduced the (random) canonical TPQ state $\vert  r_\beta \rangle = \exp \bigl(- \tfrac{\beta}{2} H\bigr) \vert  r \rangle$~\cite{sugiura:2013}.
    Most notably, the mean square error of this estimator~\autoref{eq:tpq_estimator} is exponentially small in system size at a finite free-energy density~\cite{sugiura:2013}, e.g., above the gap of a gapped system.
    Secondly, the exponential of $H$ is approximated using the Krylov basis $v_1, \ldots, v_n$ obtained from the Lanczos algorithm starting at $v_1=\vert r\rangle$, leading to~\cite{wietek:2019}
    \begin{equation}
        \langle r_\beta \vert  A \vert  r_\beta \rangle \approx e_1^\top e^{-\tfrac{\beta}{2}T_n} V_n^\dag A V_n e^{-\tfrac{\beta}{2}T_n} e_1\,,
    \end{equation}
    where $V_n = (v_1\vert  \ldots \vert  v_n)$ and $T_n$ is the tridiagonal matrix obtained after $n$ Lanczos steps. $e_1$ is the vector whose first entry is unity and whose remaining entries are zero.
    In the case $A = H^k$ diagonalizing $U_n^\dag T_n U_n = D_n$ leads to the familiar FTLM expression~\cite{schnack_magnetism_2018}
    \begin{equation}\label{eq:ftlm_estimator}
        \langle r_\beta \vert  H^k \vert  r_\beta \rangle \approx \sum_{j=1}^n  \epsilon_j^k e^{-\beta \epsilon_j} \vert \langle r \vert  u_j \rangle\vert ^2\,,
    \end{equation}
    where $u_j$ is an eigenvector of $T_n$ with eigenvalue $\epsilon_j$ and $U_n = (u_1\vert \ldots\vert u_n)$.
    We then apply \autoref{eq:tpq_estimator} and \autoref{eq:ftlm_estimator} to the block decomposition
    \begin{equation}
        \Tr[H^k e^{-\beta H}] = \sum_{\rho} \Tr_\rho[H^k_\rho e^{-\beta H_\rho}]\,,
    \end{equation}
    where $\rho$ denotes irreps labeled by $S^z_{\mathrm{tot}}$ and a space group index while $\Tr_\rho$ denotes the trace in the $\rho$ block of $H$. 
    With this setup, the Lanczos routine can be run separately on each $H_\rho$ instead of the full Hamiltonian at once.
    For the most symmetric 36-site maple-leaf lattice cluster a number of $n < 150$ Lanczos steps and $R=50$ random vectors for each block are sufficient to reduce statistical errors, which we estimate based on a jackknife analysis~\cite{efron1981jackknife} due to the ratio appearing in \autoref{eq:tpq_estimator}.

\paragraph*{(iv) Classical Monte Carlo.}
We employ O(3)-spin Monte Carlo simulations \cite{fbuessen_spin_mc} to probe the classical structure factor and specific heat between temperatures of $T=10$ down to $T=0.05$ on a lattice of size linear size $L=16$ contains $N=1536$ sites.

\begin{figure}[t]
    \centering
    \includegraphics[width=\columnwidth]{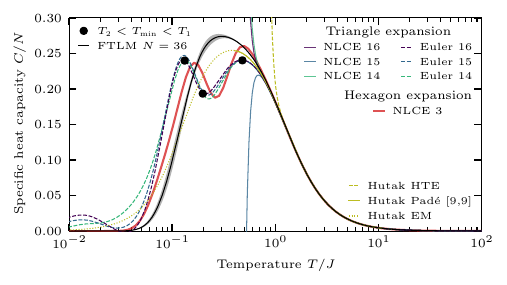}  
    \caption{Specific heat capacity as a function of temperature. We show NLCE data from the triangle-based expansion (blue/green; up to order 16) and from the hexagon-based expansion (red; order 3). FTLM data are shown for a cluster with $N = 36$ sites and periodic boundary conditions in black. The positions of the specific-heat peaks of the triangle expansion are indicated at $T_1 \approx 0.479\,J$ and $T_2 \approx 0.131\,J$, as well as the minimum at $T_{\mathrm{min}} \approx 0.198\,J$. For comparison, the high-temperature expansion, along with the Pad\'e approximation and the interpolation to zero temperature using the entropy method (EM) from Hutak~\cite{hutak_thermodynamics_2025}, is shown in yellow.} 
    \label{fig:cv}
\end{figure}

\section{Thermodynamic properties}\label{sec:thermo}
Using the numerical methods described in \autoref{sec:model_and_methds}, we compute different thermodynamic quantities, such as the energy $E$, specific heat capacity $C$, and entropy $S$:
\begin{align}
    E &= \langle H\rangle_\beta=\sum_{\varepsilon_i}\varepsilon_ie^{- \beta\varepsilon_i}/Z\\
    C &= \frac{1}{\kb T^2}\left(\langle H^2\rangle_\beta- \langle H\rangle^2_\beta\right)\\
    S &= \log(Z) -  \beta E
\end{align}
where $Z=\sum_{\varepsilon_i}e^{-\beta\varepsilon_i}$ is the partition function and the sum runs over all eigenstates. $\beta=1/(\kb T)$ is the inverse temperature. $\kb$ denotes the Boltzmann constant and is set to unity. 

\paragraph*{Specific heat.} We begin by discussing the specific heat displayed in \autoref{fig:cv}. It shows the hexagon-based (red) and triangle-based (blue/green) NLCE data, FTLM (black) for $N=36$ sites, and a comparison to Ref.~\cite{hutak_thermodynamics_2025} (yellow).
While all methods agree at high temperatures and reproduce the expected behavior, scaling as $\beta^{2}$, they exhibit qualitative differences at intermediate and low temperatures.

Hutak~\cite{hutak_thermodynamics_2025} reports a single peak at $T \approx 0.379\,J$ with $C/N \approx 0.254$. Their results are based on a conventional high-temperature expansion up to 18th order (converged down to $T \approx 1\,J$), which was combined with Pad\'e approximants to extend the analysis to temperatures around $T \approx 0.4\,J$, comparable to the first peak. The low-temperature behavior was then estimated using the entropy method, which interpolates based on the high-temperature data, the ground-state energy, and assumptions about low-energy excitations. They assume that the low-temperature physics is dominated by antiferromagnetic spin-wave excitations, leading to $C \propto T^2$ as $T \rightarrow 0$~\cite{minoru_spin_1989}.

\begin{figure}[t]
    \includegraphics[width=\columnwidth]{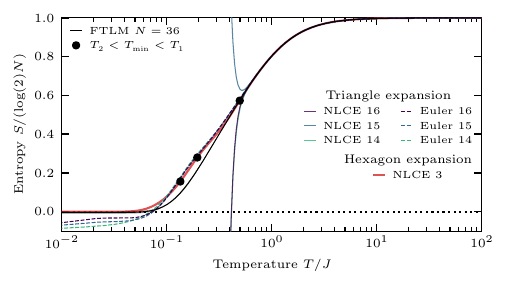}
    \caption{Entropy as a function of temperature. We include NLCE results from the triangle-based expansion (blue/green; up to order 16) and from the hexagon-based expansion (red; order 3). FTLM data for a cluster with $N = 36$ sites and periodic boundary conditions are also shown in black. Markers indicate the positions of the two peaks and the minimum in the specific heat capacity, as presented in \autoref{fig:cv}.}
    \label{fig:entropy}
\end{figure}

This observation is consistent with the FTLM calculation, which shows a single, comparatively broad peak extending over nearly two decades in temperature. Such a broad feature in the specific heat may indicate the presence of multiple relevant energy scales in the system.

The NLCE algorithm is fundamentally a high-temperature method, and observables are expected to diverge at intermediate and low temperatures once the correlation length exceeds the size of the largest cluster included in the expansion. In this case, the combination of resummed weights and exponentially growing embedding factors causes the series to alternate between positive and negative values across consecutive orders, signaling the breakdown of the expansion. The Euler series acceleration provides a resummation of these alternating divergences and extends the accessible temperature range somewhat, but it is similarly expected to encounter divergences at sufficiently low temperatures. 

Remarkably, we observe a distinct scenario here. While the bare NLCE data for the triangular expansion exhibit the expected alternating divergence at $T\approx 0.6\,J$ the Euler resummation successfully removes this divergence and yields physically reasonable results down to much lower temperatures. Although consecutive orders do not match perfectly, their overall shape appears consistent. In contrast, for the hexagon-based expansion, the bare NLCE data entirely lack the characteristic divergence. This absence of divergence indicates that the data provide a meaningful description of the thermodynamics. Other expansion units do display the expected divergent behavior. 

Both expansions suggest the existence of a second peak at around $T_2\approx 0.13\,J$, in addition to the first peak at $T_1\approx 0.48\,J$. The appearance of a second peak indicates the presence of a second energy scale in the system, corresponding to distinct excitations that freeze out at different temperatures. This effect is commonly observed in frustrated magnets~\cite{zhongling_strained_2024,popp_origin_2025}. Evidence for such a two-peak structure has also been reported for the related triangular lattice at comparable temperatures~\cite{chen_two-temperature_2019, ulaga_finite_2024}. In contrast, data for the kagome lattice only suggest a broad shoulder instead of a distinct second peak~\cite{bernu_effect_2020, schnack_magnetism_2018}.

\begin{figure}[t]
    \centering
    \includegraphics[width=\columnwidth]{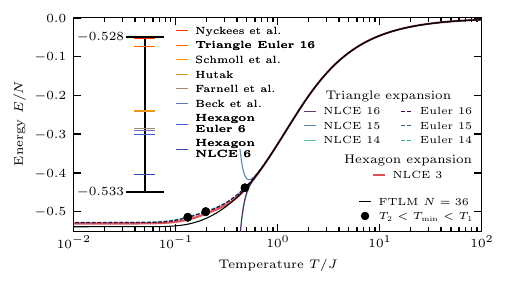}
    \caption{Energy as a function of temperature. It shows the NLCE data from the triangle-based (blue/green; up to order 16) and hexagon-based (red; order 3) expansions are shown, together with FTLM results for a periodic $N = 36$ cluster (black). The positions of the specific-heat peaks from \autoref{fig:cv} are marked with black dots. The inset in the energy panel displays the ground-state energy estimates. It contains the literature results shown in \autoref{tab:energies} from Refs.~\cite{nyckees_tensor_2025,schmoll_bathing_2025,hutak_thermodynamics_2025,farnell:2011, beck:2024} and NLCE estimates from the triangle expansion (order 16) and the hexagon expansion (order 6), shown in bold fonts.} 
    \label{fig:E}
\end{figure}

\begin{table}[t]
    \centering
    \begin{tabular}{l l}
        Method/Ansatz &  $E_0$/ $N J$ \\
        \hline
        \textbf{Isolated hexagons} & \textbf{-0.4671}\\
        Linear spin wave theory~\cite{schmalfuss:2002} & -0.5121574375 \\
        Neural quantum states, Ref.~\cite{beck:2024} & -0.523571\\
        iPEPS, Ref.~\cite{nyckees_tensor_2025} &  -0.528052 \\ 
        \textbf{Triangular Euler 16} & \textbf{-0.5285(3)} \\ 
        iPEPS, Ref.~\cite{schmoll_bathing_2025}     & -0.530359 \\
        Entropy method~\cite{hutak_thermodynamics_2025} & -0.5304(2)\\
        Coupled cluster method~\cite{farnell:2011}& -0.53094 \\
        iDMRG, Ref.~\cite{beck:2024} & -0.531003 \\
        \textbf{Hexagon Euler 6} & \textbf{-0.5311(9)} \\
        \textbf{Hexagon NLCE 6} & \textbf{-0.532(2)} \\ 
        \textbf{Exact diagonalization $\mathbf{N=36}$} & \textbf{-0.538972}
    \end{tabular}
    \caption{Estimates for the ground-state energy per spin from the literature and this work. The comparison includes an infinite Projected Entangled Pair States (iPEPS)~\cite{jordan_classical_2008} study~\cite{schmoll_bathing_2025,nyckees_tensor_2025}, coupled cluster calculation~\cite{farnell:2011}, linear spin wave simulations~\cite{schmalfuss:2002}, neural quantum states~\cite{beck:2024}, infinite density matrix renormalization group~\cite{beck:2024}, and an estimate through the entropy method~\cite{hutak_thermodynamics_2025}.
    The results from this work, highlighted in bold characters, comprise NLCE data from the triangular expansion (Euler 16), and both bare (NLCE 6) and Euler-resummed data (Euler 6) from the hexagonal expansion at order six, as well as the exact calculation for $N = 36$ sites with periodic boundaries. The latter was also computed in Ref.~\cite{schmalfuss:2002}. The errors of the NLCE estimates are defined as the difference between the highest and the second-highest expansion order. A visual representation is provided in \autoref{fig:E}.}
    \label{tab:energies}
\end{table}

\paragraph*{Entropy.} We observe a similar convergence for the entropy as a function of temperature, as shown in \autoref{fig:entropy}. It includes results from the triangular and hexagonal expansions, as well as FTLM data for the $N=36$ cluster. All methods agree at high and intermediate temperatures for $T \gtrsim T_1$. We find that the finite-size cluster releases its entropy faster than predicted by the cluster expansions, and we do not observe any indication of an intermediate plateau.

Similarly, the bare NLCE expansion for the triangular expansion diverges, but the Euler resummation yields physically reasonable results for $T \gtrsim 0.1\,J$, while the entropy becomes negative below this temperature. The curves approach zero from below as the order increases, indicating systematic improvement, and the trend suggests that the entropy would likely reach zero with the inclusion of a few additional orders. Remarkably, the third-order hexagon expansion correctly predicts zero entropy for $T \lesssim 0.05\,J$, indicating that these results are physically meaningful and supporting our interpretation of a second peak in the heat capacity.

\paragraph*{Energy.} The energy as a function of temperature is displayed in \autoref{fig:E}. In the hexagon expansion, the bare NLCE data remain finite and physically reasonable. For the triangular expansion, the bare NLCE curves diverge around $T \approx 0.6 \,J$, while the Euler resummation successfully produces realistic results. At high to intermediate temperatures, both expansions are in agreement with the FTLM results, which yield a slightly lower ground-state energy. Finite clusters with periodic boundary conditions tend to underestimate the ground-state energy compared to the thermodynamic limit.

The convergence of the NLCE further enables an estimate of the ground-state energy. The inset in \autoref{fig:E} and \autoref{tab:energies} compares the NLCE estimates with previously reported results. Note that, at zero temperature, the NLCE algorithm can be extended to higher orders by employing the Lanczos method to reach larger system sizes. In particular, we have computed the ground-state energy for all 147 topologically inequivalent clusters within the hexagon expansion up to sixth order (clusters of size $N=36$)~\cite{DanceQ,DanceQ_code}. The ground-state energy remains stable across multiple expansion orders, indicating that the algorithm has indeed converged in the zero-temperature limit. We remark that this value is not variational and therefore does not represent a lower bound.

The reason for the convergence is that the nearest-neighbour correlations in the ground state of each topologically invariant cluster built from hexagons remain remarkably stable. The nearest-neighbour correlation directly encodes the energy, and it is unusual that these correlations do not vary significantly across clusters. One would typically expect open boundaries to induce substantial changes in the ground-state structure. Instead, we find a consistent pattern in which the hexagonal, triangular, and diagonal bonds each exhibit nearly identical correlations, independent of the cluster geometry. We have shown the nearest-neighbour correlations for all hexagonal clusters up to order four in \autoref{app:NLCE}.

\begin{figure}[t]
    \includegraphics[width=\columnwidth]{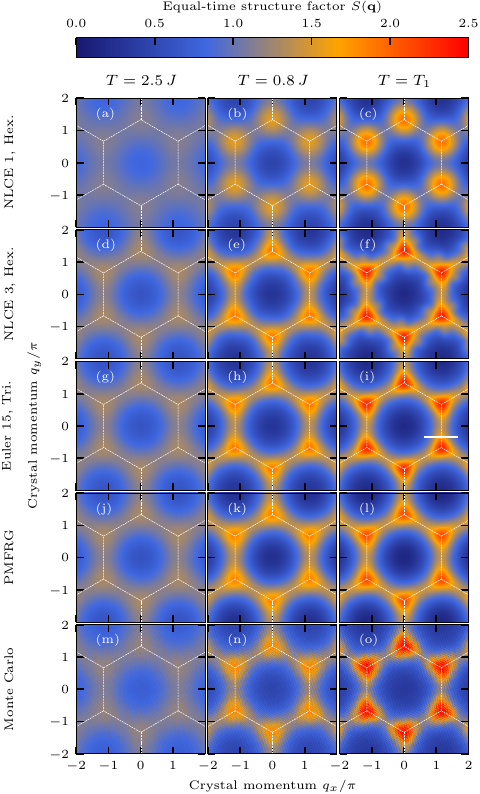}
    \caption{Equal-time spin structure factor computed using different numerical methods and at various temperatures.
    The first, second, and third columns correspond to $T = 2.5\,J$, $T=0.8\,J$, and the Schottky peak at $T = T_1 \approx 0.47\,J$, respectively. Panels (a–c) show the bare NLCE first-order hexagon expansion (decoupled-hexagons limit), panels (d–f) display the third-order hexagon expansion, and panels (g–i) present the 15th-order triangular expansion using Euler resummation, panels (j-l) show PMFRG results, and panels (m-o) refer to the classical Monte Carlo Data results. For details on how we compare with classical Monte Carlo data, see \autoref{app:MC}.}
    \label{fig:Sq_highT}
\end{figure}

\section{Structure factor}\label{sec:StructureFactor}

We now turn to the equal-time spin structure factor to resolve the spin correlation at finite and zero temperature. It is defined as
\begin{align}
    S(\vec{q}) = \frac{4}{3N} \sum_{i,j} e^{i\vec{q}\cdot\left(\vec{r}_i-\vec{r}_j\right)}\left\langle\vec{S}_i\cdot \vec{S}_j\right\rangle_\beta\,.
\end{align}
The prefactor ${4}/{3}$ normalizes to unity in the infinite-temperature limit.

\begin{figure}[t]
    \includegraphics[width=\columnwidth]{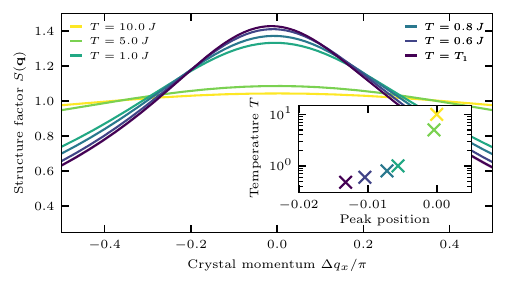}
    \caption{Equal-time structure factor along the line cut $q_y/\pi = -1/3$ and $q_x/\pi \in [2/\sqrt 3 -1/2, 2/\sqrt 3 +1/2]$, as highlighted in \figref{fig:Sq_highT}{i}, shown for different temperatures. The $x$-axis is centered around $\Delta q_x/\pi = q_x/\pi - 2/\sqrt 3$. The inset displays the temperature dependence of the peak position.}
    \label{fig:Sq_twist}
\end{figure}

We begin by discussing the high- to intermediate-temperature regime shown in \autoref{fig:Sq_highT}, where we compare different numerical approaches outlined in \autoref{sec:model_and_methds}. The columns (from left to right) show results for $T = 2.5\,J$, $T = 0.8\,J$, and $T = T_1$, where $T_1$ denotes the position of the high-temperature peak.

 The first row corresponds to a first-order hexagon expansion, which reduces to a set of uncoupled, independent hexagons.
The second row shows results from the third-order hexagon expansion (bare NLCE). The third row presents the 15th-order triangular expansion obtained using the Euler resummation scheme. PMFRG results are displayed in the fourth row, and the fifth row shows the classical Monte Carlo simulations, where we have adjusted the temperature and intensity as detailed in \autoref{app:MC}.

At the highest temperature (first column), $T = 2.5\,J$, all methods display a consistent pattern that reflects the geometry of the underlying triangular lattice. We observe enhanced intensity at the $K$ points located at $\left(0,\pm \frac{4\pi}{3}\right)$ and $\left(\pm \frac{2\pi}{\sqrt{3}}, \pm\frac{2\pi}{3}\right)$. The decoupled-hexagon ansatz, shown in \figref{fig:Sq_highT}{a}, exhibits a slight deviation characterized by a reduced overall intensity, stemming from the reduced coordination number of this ansatz.

At intermediate temperature (second column, $T = 0.8,J$), the decoupled-hexagon ansatz exhibits qualitatively different behavior, as the intensity between the $K$ points is suppressed in \figref{fig:Sq_highT}{b}. In contrast, the other approaches show a continued enhancement of the $K$-point intensity and remain in good mutual agreement. We observe an intensified signal at the $K$ points, forming a triangular plateau.

At the first peak (third column, $T = T_1 \approx 0.47\,J$), the triangular plateaus around the $K$ points become more pronounced. However, rather than developing into sharp features, they remain broad and flat. The third-order hexagon expansion, shown in \figref{fig:Sq_highT}{f}, displays the onset of a less smooth texture, which we attribute to insufficient convergence at these intermediate temperatures. The other methods remain largely consistent. We just observe a minor discrepancy in the PFMRG intensities, which is analyzed in \autoref{app:PMFRG}.
The 15th-order triangular expansion in \figref{fig:Sq_highT}{i} is well converged, as it agrees with lower-order results (see \autoref{app:NLCE}) and provides a meaningful basis for comparison to experiments.

\begin{figure}[t]
    \includegraphics[width=\columnwidth]{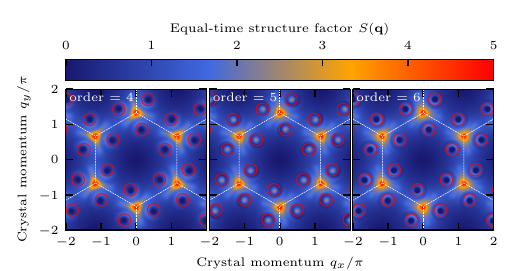}
    \caption{Equal-time structure factor of the hexagon expansion at $T = 0$ for orders four, five, and six. The intensity and shape of the triangular plateaus around the $K$ points, including their characteristic twist, remain stable across orders. In contrast, the emergent satellite peaks, highlighted in red, at the flanks alternate in sign, indicating a lack of convergence. They are located at the $2M$ point and associated with N\'eel order~\cite{gresista_unconventional_2025}.}
    \label{fig:Sq_hexagons}
\end{figure}

At temperatures comparable to the occurrence of the first peak, we observe the emergence of an additional feature in the structure factor that is absent at high temperatures. The triangular plateaus develop a subtle twist, where alternating triangles appear slightly rotated clockwise and counterclockwise.
This twisting pattern is consistently observed across all numerical methods except for the decoupled-hexagon ansatz. It becomes increasingly pronounced at lower temperatures.
Notably, this twist is consistent with the absence of reflection symmetry in the lattice and, therefore, not present in the triangular or kagome lattice~\cite{mueller_thermodynamics_2018,chen_two-temperature_2019}.
A similar effect can be seen in other studies at zero temperature~\cite{beck:2024, sonnenschein_candidate_2024}.

We quantify the twist in \autoref{fig:Sq_twist}, where we plot the intensity of $S(q_x, q_y)$ along the line cut indicated in \figref{fig:Sq_highT}{i} for a fixed $q_y/\pi = -1/3$, as a function of $q_x$ at different temperatures. The horizontal axis shows $\Delta q_x$, defined relative to the high-symmetry line connecting the two vertical $K$ points. While the peak is symmetric in the high-temperature limit, it shifts toward negative $\Delta q_x$ for lower temperatures. The inset shows the position of this peak, providing a quantitative measure of the observed twist.

The convergence of the NLCE algorithm at $T = 0$ enables the extension of the hexagon-based expansion to higher orders, incorporating all 147 topologically inequivalent clusters containing $N = 36$ sites at sixth order. The corresponding structure factors for orders four, five, and six are shown in \autoref{fig:Sq_hexagons} (bare NLCE data). The triangular plateaus around the $K$ points remain stable in both shape and intensity, and the characteristic twist persists across orders. Similar to the finite-temperature results of the third-order expansion shown in \figref{fig:Sq_highT}{d–f}, we observe the emergence of satellite peaks at the sides of the triangular plateaus. These peaks alternate in sign between consecutive orders, indicating a lack of convergence. They are located at the $2M$ point (see Ref.~\cite{gresista_unconventional_2025}) and associated with N\'eel order. We therefore conclude that the convergence of the expansion is $\mathbf{q}$-dependent.
The finite-size calculation of the $N=36$ cluster with periodic boundary condition also indicates the presence of the twist at $T=0$, which we have shown in~\autoref{app:ED-structure-Factor}.

Ref.~\cite{gresista_unconventional_2025} presents exemplary structure factors obtained using the pseudo-fermion functional renormalization group across a phase diagram with ferromagnetic diagonal and triangular bonds. The zero-temperature calculation shown in \autoref{fig:Sq_hexagons} resembles the paramagnetic pattern (PM E), which lies near the canted $120^\circ$ antiferromagnetic order. It also shares features with the background of the $120^\circ$ state, although we do not observe a sharp response at the $K$ points in our NLCE calculation.

\begin{figure}[t]
    \includegraphics[width=\columnwidth]{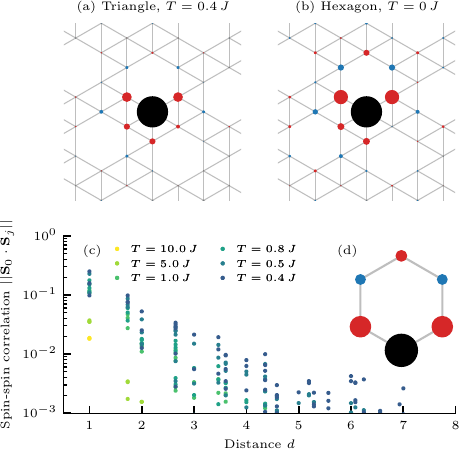}
    \caption{Real-space correlations. Panel (a) presents data from the 15th-order triangular expansion at finite temperature $T = 0.4\,J$, obtained using Euler resummation. Panel (b) shows results from the 6th-order NLCE hexagon expansion at zero temperature. Panel (d) displays data for an isolated hexagon. The marker size represents the absolute value of $\langle \vec{S}_j \cdot \vec{S}_0 \rangle_\beta$, where $\vec{S}_0$ denotes the central spin (shown in black) with a reference value of $0.75$. The color indicates the sign of the correlation: red refers to negative and blue refers to positive. (c) Absolute value of the spin-spin correlation as a function of distance for different temperatures, computed the 15th-order triangular expansion and the Euler resummation.}
    \label{fig:Sq_points}
\end{figure}
\paragraph*{Real-space correlation.}
Lastly, we discuss the real-space correlations, which are directly computed by the NLCE algorithm and is used to compute the structure factors. \figref{fig:Sq_points}{a,b} show a subsection of the lattice, displaying the correlations of the central spin $\vec{S}_0$ (marked in black) with all other sites. The marker size encodes the magnitude of the correlation $\langle \vec{S}_0 \cdot \vec{S}_j \rangle_\beta$, while the color indicates its sign (red for negative, blue for positive). The central spin’s marker size serves as a reference, corresponding to $\langle \vec{S}_0 \cdot \vec{S}_0 \rangle_\beta = 0.75$. 
 
\figref{fig:Sq_points}{a} presents the triangular expansion (Euler resummation, 15th order) at finite temperature $T = 0.4\,J$, and \figref{fig:Sq_points}{b} shows the sixth-order hexagon expansion (bare NLCE) at $T = 0$. The full cluster accessible at sixth order in the hexagon expansion, comprising 546 sites, is shown in \autoref{app:NLCE}. Both display a consistent spatial pattern, indicative of a short-range correlated state with negligible correlations beyond third-nearest neighbors. Particularly noticeable is the correlation across individual hexagons, resembling the pattern expected for decoupled hexagons at zero temperature, which is shown in \figref{fig:Sq_points}{d}.

\begin{figure}[t]
    \includegraphics[width=\columnwidth]{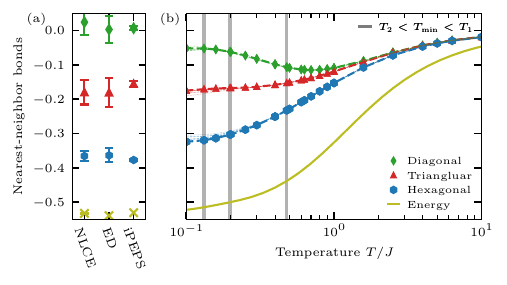}
    \caption{Nearest-neighbor correlations at zero temperature (a) and as a function of temperature (b). The green, red, and blue curves represent the diagonal, triangular, and hexagonal bonds, respectively. Panel (a) shows estimates from the 6th-order NLCE hexagon expansion, ED for the $N=36$ site cluster, and iPEPS data from Schmoll et al.~\cite{schmoll_bathing_2025}. The error bars for the NLCE data are defined as the difference from the 5th order, while those for ED and iPEPS correspond to the standard deviation within each group of bond types. For instance, there are multiple diagonal bonds within the cluster that yield slightly different values. Panel (b) displays the 15th-order Euler-resummed data from the triangular expansion; lower orders are shown with thinner lines for comparison. The yellow curve denotes the total energy obtained by summing all correlations weighted by their respective multiplicities.}
    \label{fig:real space_nn}
\end{figure}

\figref{fig:Sq_points}{c} quantifies these short-range correlations further. It shows the absolute value of the spin-spin correlation versus distance $d$ (in lattice units) as a function of temperature, computed from the triangular expansion up to order 15 (Euler resummation) at $T = 0.4\,J$. Notably, this temperature already lies close to the ground-state regime, as the energy $E(T = 0.4\,J) \approx 0.5\,J$ is near the estimated ground-state energy (see \autoref{tab:energies}).

The comparison between the finite- and zero-temperature data in \figref{fig:Sq_points}{a,b} highlights a key qualitative difference. At finite temperature, the correlations along the diagonal bonds are negative and comparable in magnitude to those on the other nearest-neighbor bonds, whereas in the $T=0$ data these correlations are nearly absent. This behavior is qualitatively shown in \figref{fig:real space_nn}{b}, which displays the temperature dependence of the correlation on the three inequivalent bond types: green for diagonal bonds, red for triangular bonds, and blue for hexagonal bonds (as defined in \autoref{fig:maple_leaf}). The respective solid lines denote the 15th-order Euler-resummed results, while the thinner dashed lines indicate lower-order approximations, illustrating both the evolution between consecutive orders and the level of convergence. The data clearly demonstrate that the diagonal-bond correlations grow in magnitude up to $T \approx 0.7\,J$ before weakening again as the temperature decreases, producing a distinctly non-monotonic temperature dependence. The nearest-neighbor correlators appear converged down to $T \approx 0.2\,J$, indicating that this feature is robust and well resolved.

\figref{fig:real space_nn}{a} shows the corresponding zero-temperature estimates for comparison. It includes the result from the sixth-order hexagon expansion (first column), ED data for the $N=36$ cluster (second column), and the iPEPS simulation from Ref.~\cite{schmoll_bathing_2025} (third column). The error bars of the hexagon expansion are defined as the difference from the fifth-order result. The error bar of the ED data arises from the non-symmetric cluster shape, which causes equivalent bonds (e.g., diagonal) to exhibit different correlations. The error is defined as the standard deviation of all equivalent bonds. Overall, we find that all ground-state estimates are in good agreement. They place the diagonal-bond correlator close to zero.
The plot further includes the corresponding ground-state energy estimates in yellow.

By comparing the zero-temperature data in \figref{fig:real space_nn}{a} with the finite-temperature data in \figref{fig:real space_nn}{b}, we observe that the triangular-bond correlators (red) already show very good agreement between the two. In contrast, the diagonal-bond (green) and hexagonal-bond (blue) correlators still exhibit noticeable deviations. However, the flow with increasing expansion order (dashed lines in \figref{fig:real space_nn}{b}) gradually approaches the corresponding ground-state estimates. Notably, although the nearest-neighbor correlation disagree with the zero-temperature results for the diagonal and hexagonal bonds, the total energy nevertheless provides a good approximation to the ground-state value, shown in yellow. This arises because the deviations between the hexagonal and triangular finite-temperature estimates largely cancel once the correct multiplicities are taken into account (there are twice as many hexagonal as diagonal bonds).

\section{Discussion}
We study the Heisenberg antiferromagnetic maple-leaf lattice using a variety of numerical methods. We present novel finite-temperature results and provide new insights into its ground-state properties. Our analysis primarily relies on the NLCE algorithm, which exhibits rather unconventional convergence, allowing us to access a non-trivial temperature regime and yielding reliable data even at zero temperature. The finite-temperature results presented here lay important groundwork for forthcoming experimental studies.

Both series expansions discussed here indicate the presence of a two-peak structure in the specific heat. Although this differs from the single broad peak reported in Ref.~\cite{hutak_thermodynamics_2025} and from our FTLM simulations, it is consistent with observations for the closely related triangular lattice~\cite{chen_two-temperature_2019, ulaga_finite_2024}, where the emergence of two characteristic temperature scales has likewise been proposed. For the triangular lattice, Ref.~\cite{chen_two-temperature_2019} suggested that the higher-energy scale is associated with magnon excitations, whereas the lower-energy scale may correspond to the so-called ``roton-like'' excitations at the $M$ point of the Brillouin zone.
It is also worth noting that the kagome lattice exhibits a broad shoulder, rather than a single Schottky peak, in its specific heat capacity~\cite{bernu_effect_2020, schnack_magnetism_2018}.

We further investigate the equal-time structure factor, which is directly accessible in neutron-scattering experiments. At high temperatures, it closely resembles the results obtained for the triangular and kagome lattices~\cite{mueller_thermodynamics_2018, chen_two-temperature_2019}. However, upon approaching intermediate temperatures around the first specific-heat peak, an emergent twist becomes apparent, which is possible due the absence of reflection symmetry in the maple-leaf lattice. In contrast, the related triangular-based lattices that preserve reflection symmetry do not exhibit this feature.

The convergence of the NLCE algorithm extends down to zero temperature, enabling estimates of both the ground-state energy and the structure factor. This behavior is uncommon and suggests the presence of an unordered ground state with short-range correlations since the algorithm performs poorly when approaching states with a long correlation length. We emphasize that NLCE is inherently a high-temperature expansion, and its results are not of variational nature. Nevertheless, the agreement with other numerical methods --- particularly for the ground-state energy --- suggests that the obtained physical insights are reliable. The resulting structure factor indicates that the twist persists, and we find no evidence of magnetic order.

A similar zero-temperature convergence within the NLCE framework was previously reported in Ref.~\cite{schaefer_abundance_2023} for an equivalent setup. In that work, expansions based on hexagons (for the pyrochlore and ruby lattices) and on squares (for the checkerboard lattice) also exhibited convergence of the ground-state energy. The authors attributed this behavior to the presence of a relatively simple valence-bond crystal composed of weakly coupled, unfrustrated loops (hexagons or squares). These unfrustrated motifs are connected through tetrahedral units, which lead to a weak coupling or ``dressing''. The corresponding state was characterized using a straightforward variational ansatz,
\begin{align}
    \ket{\Psi_\alpha} \propto e^{-\alpha V}\ket{\hexagon},\label{eq:hexagon_pyro}
\end{align}
where $\ket{\hexagon}$ denotes a product state in which each (non-overlapping) hexagon is in its local ground state.
$V$ represents the inter-hexagon coupling (mediated through the tetrahedra in Ref.~\cite{schaefer_abundance_2023}), and $\alpha$ is a variational parameter.

For the maple-leaf lattice, an analogous construction can be employed. Here, the inter-hexagon coupling includes two inequivalent bonds (green and red in \autoref{fig:maple_leaf}). These two couplings, denoted $V_g$ and $V_r$, can be incorporated into the following variational wavefunction:
\begin{align}
    \ket{\Psi_{(\alpha_g,\alpha_r)}} \propto e^{-\alpha_g V_g-\alpha_r V_r}\ket{\hexagon}\,.\label{eq:hexagon}
\end{align}
It is also worth mentioning that the algorithm employed in Ref.~\cite{schaefer_abundance_2023} exhibited unphysical behavior at finite temperatures, such as a non-monotonic temperature dependence of the energy.
In contrast, no such artifacts are observed in our results.

Notably, the state described above is paramagnetic, as it does not break any lattice symmetries. Its excitations are localized triplets that do not disperse. These localized excitations are separated from the continuum, as illustrated in Fig. 4 of Ref.~\cite{schaefer_abundance_2023}.

Large-scale iPEPS calculations~\cite{schmoll_bathing_2025} further support the existence of a paramagnetic ground state in this regime. A hexagon singlet was also proposed away from the isotropic point~\cite{ghosh_where_2024,ghosh:2024b}. Our data indicate that this state occupies a broader portion of parameter space than previously assumed. However, we also note that ongoing studies~\cite{ebert:2026} based on finite-size ED simulations suggest the presence of magnetic order at the isotropic point. This ordered state, however, lies very close to the paramagnetic hexagonal state that emerges in the limit of dominant hexagonal bonds. 

All data used to generate the plots is available on Zenodo~\cite{scheafer_2026_18222467}.
\begin{acknowledgments}
This work is dedicated to the memory of Johannes Richter. As a pioneer in frustrated magnetism and numerical techniques, he also inspired a younger generation by developing many of the methods employed in this study. We are grateful for his contributions, many of which are of direct relevance and significance to this study.

We acknowledge valuable discussions with Yasir Iqbal and Nils Niggemann. We further thank Philipp Schmoll, Taras Hutak, and Pratyay Ghosh for kindly sharing their data with us. We also acknowledge the mathematical construction by Hamid Naderi Yeganeh that was used to draw the maple leaf in \autoref{fig:maple_leaf}. Simulations were performed using \texttt{SpinMC.jl}~\cite{fbuessen_spin_mc}, \texttt{PMFRG.jl}~\cite{niggemann2021frustrated,niggemann2022quantitative,niggemann_PMFRG}, \texttt{XDiag}~\cite{wietek:2025:xdiag}, and \texttt{DanceQ}~\cite{DanceQ,DanceQ_code}.  A.W. acknowledges support from the German Research Foundation (DFG) through the Emmy Noether program (Grant No.~509755282). N.H. acknowledges support from the German Research Foundation (DFG), within Project-ID 277101999 CRC 183 (Project A04).
D.J.L. acknowledges support by the DFG through the cluster of excellence ML4Q (EXC 2004, project-id 390534769), CRC1639 NuMeriQS (project-id  511713970) and CRC TR185 OSCAR (project-id 277625399). 
\end{acknowledgments}

\bibliography{ref}

\clearpage

\appendix

\section{Numerical linked cluster expansion}\label{app:NLCE}
This appendix presents additional NLCE results, illustrates the convergence behaviour, and provides further technical details.

\subsection*{Multiplicities} 
We show the number of connected and topologically inequivalent clusters per order in \autoref{tab:multi}. For the hexagon expansion, the cluster size at a given order is simply obtained by multiplying the order by six. In contrast, the cluster sizes in the triangle expansion vary because the triangles share edges and sites. The largest cluster we include at 16th order contains sixteen sites.

\begin{table}[h]
\centering
\begin{tabular}{l | |  c c |  c c}
 & \multicolumn{2}{c| }{Hexagon} & \multicolumn{2}{c}{Triangle}  \\
order & \# Con. & \# Top. & \# Con. & \# Top. \\\hline
1 & 1 & 1 & 8 & 1 \\
2 & 3 & 1 & 9 & 1 \\
3 & 11 & 3 & 12 & 1 \\
4 &  44 & 10 & 17 & 3 \\
5 &  186 & 33 & 24 & 3 \\
6 &  814 & 147 & 36& 6 \\
7 &  3652 & 620 & 56 & 9 \\
8 &  16689 & 2821 & 87 & 15 \\
9 &  77359 & 12942 & 138 & 22 \\
10 &  -- & -- & 224 & 39 \\
11 &  -- & -- & 366 & 60 \\
12 &  -- & -- &  603 & 103 \\
13 &  -- & -- & 1006 & 168 \\
14 &  -- & -- &  1692 & 287 \\
15 &  -- & -- & 2862 & 476 \\
16 &  -- & -- & 4875 & 823 \\
\end{tabular}
\caption{Number of connected (\# Con.) and topologically inequivalent (\# Top.) clusters for the hexagon and edge-sharing triangle expansions, by order.}\label{tab:multi}
\end{table}
\begin{figure}[t]
    \centering
    \includegraphics[width=\columnwidth]{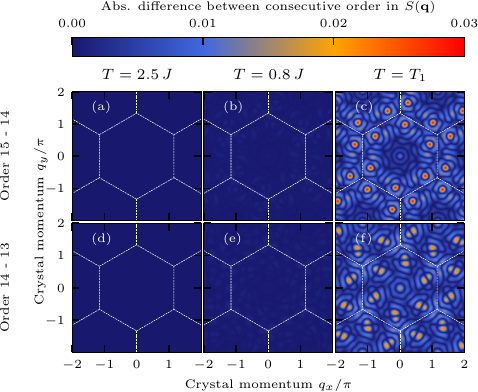}
    \caption{Difference between consecutive orders of the triangular expansion for the same temperature $T=2.5\,J,0.8\,J$ and $T=T_1$ as used in \autoref{fig:Sq_highT} in the main text. Panels (a–c) show the absolute value of the difference between the $15$th and $14$th orders, while panels (d–f) display the difference between the $14$th and $13$th orders.} 
    \label{fig:Sq_error}
\end{figure}
\subsection*{Structure factor}
We present the convergence of the structure factor for the Euler–resummed data of the triangular expansion in \autoref{fig:Sq_error}. Panels (a–c) show the absolute difference between the $15$th and $14$th orders, while panels (d–f) show the difference between the $14$th and $13$th orders. The columns correspond to the same temperatures as in \autoref{fig:Sq_highT}. Similar to the hexagon expansion shown in \autoref{fig:Sq_hexagons} at zero temperature, we observe the emergence of satellite peaks at $T = T_1$ near the $K$ point, which we attribute to insufficient convergence. These points are associated with N\'eel order~\cite{gresista_unconventional_2025}.
\subsection*{Real-space correlation}

\begin{figure*}[t]
    \centering
    \includegraphics[width=\columnwidth]{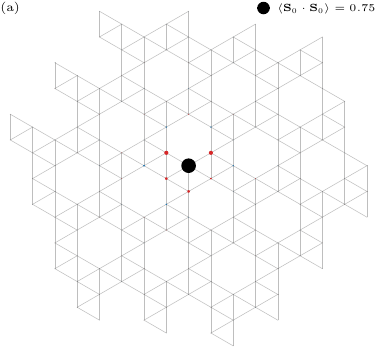}
    \hfill
    \includegraphics[width=\columnwidth]{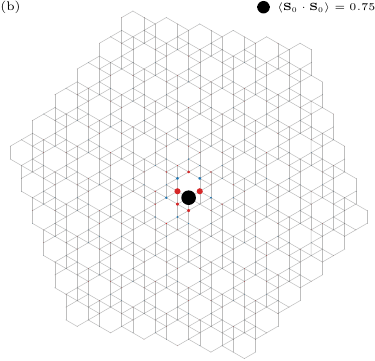}
    \caption{Real-space correlations obtained from the triangular expansion (Euler resummation) up to order $15$ at a temperature $T = 0.4\,J$ are shown in panel (a). Panel (b) shows the corresponding real-space correlations from the hexagon expansion (bare NLCE) up to sixth order at $T = 0$. The sixth order includes all clusters required to capture correlations across a total of 546 sites.} 
    \label{fig:real_space_appendix}
\end{figure*}

Within NLCE, the structure factor is obtained from the Fourier transformation of the real-space correlations, which are extracted from all generated connected clusters. For additional illustration, we show the corresponding real-space correlations in \autoref{fig:real_space_appendix}. The figure displays the correlations between the spin marked in black and all surrounding sites. Red indicates negative correlations and blue indicates positive ones, where the marker size is proportional to the correlation strength. For reference, the black marker refers to the value $\vec{S}_0 \cdot \vec{S}_0 = 0.75$.

Panel (a) shows the Euler–resummed data of the triangular expansion at $T = 0.4\,J$ for order 15. Panel (b) shows the sixth–order hexagon expansion, which contains all 147 topologically inequivalent clusters (814 translationally distinct clusters) and includes correlations across a total of 546 sites. In both cases, we find that the correlation extends only to a few neighbors, supporting the scenario of a paramagnetic state with short-range correlation.

\subsection*{Zero-temperature limit}
The convergence of the expansions at zero temperature allows us to estimate the ground-state energies. In \autoref{tab:gs_hexagon} and \autoref{tab:gs_triangle}, we show the values obtained at each order for the hexagon-based and triangle-based expansions, respectively. For the triangle expansion, only the Euler-resummed results converge; therefore, we omit the bare NLCE data.

\begin{table}[h]
\centering
\begin{tabular}{c c|  c c}
 \multicolumn{2}{c |}{NLCE} & \multicolumn{2}{c}{Euler}  \\
Order & $E_0/N$ in $J$  & Order & $E_0/N$ in $J$\\
\hline
1 & -0.4671292729553332 & -- &-- \\
2 & -0.5435706351900584 & -- & -- \\
3 & -0.5304718237518301 &3 & -0.5211855917718200\\
4 & -0.5322554996464219 &4 & -0.5276890186784275\\
5 & -0.5303844459765095 &5 & -0.5302280750492965\\
6 & -0.5324194928591225  &6 & -0.5311436689259317\\
\hline
\end{tabular}
\caption{Ground-state energies at different orders using bare NLCE (left) and Euler resummation (right) for the hexagon expansion. The Euler resummation scheme involves lower-order contributions; hence, it is not applicable for orders one and two.}\label{tab:gs_hexagon}
\end{table}

\begin{table}[t]
\centering
\begin{tabular}{ c c |  c c}
Order & $E_0/N$ in $J$ & Order & $E_0/N$ in $J$ \\\hline
3 & -0.8848033905932743 & 10 & -0.5360332092344035 \\
4 & -0.7381473012107141 & 11 &  -0.5305607132480245 \\
5 & -0.6361899307289933 & 12 & -0.5279783501768268  \\
6 &  -0.5907746777868229 & 13 & -0.5276255333285051 \\
7 & -0.5672337433600627 & 14 & -0.5283138199679275\\
8 &  -0.5542641225018704 & 15 & -0.5288440985857853 \\
9 &  -0.5440520614802178 & 16 & -0.5285335519226874 \\
\end{tabular}
\caption{Ground-state energy per order in the triangular expansion using the Euler resummation.}\label{tab:gs_triangle}
\end{table}

\begin{figure}[t]
    \includegraphics[width=0.49\columnwidth]{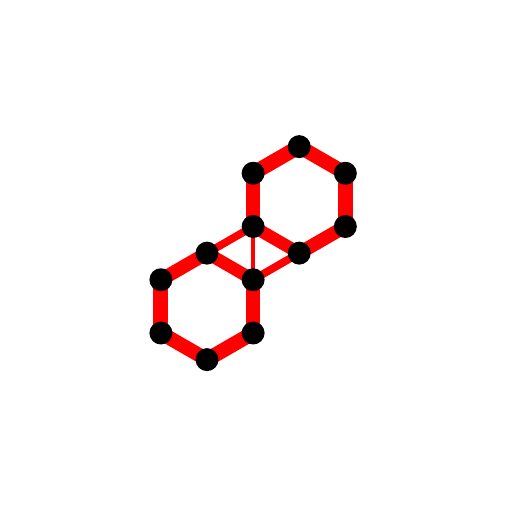}
    \includegraphics[width=0.49\columnwidth]{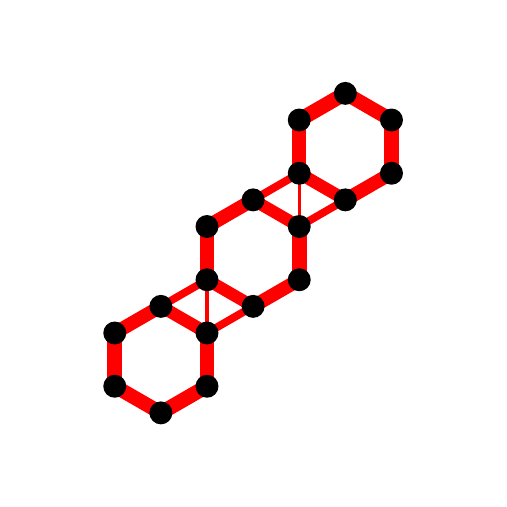}
    \includegraphics[width=0.49\columnwidth]{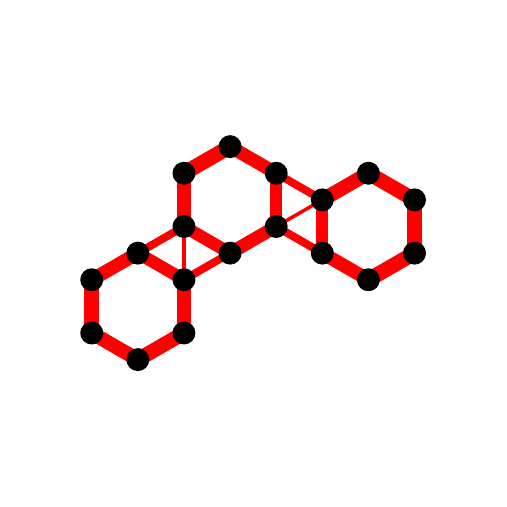}
    \includegraphics[width=0.49\columnwidth]{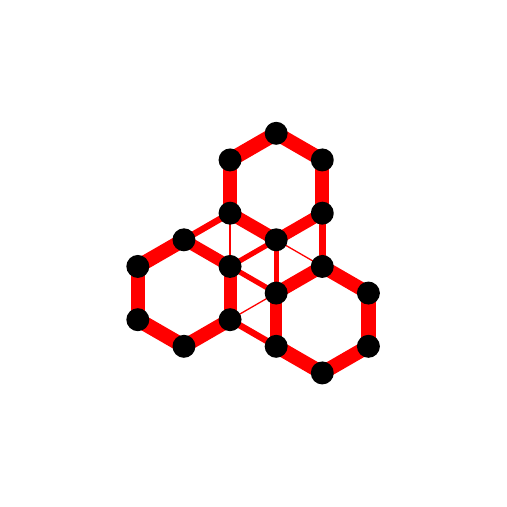}
    \caption{Topologically inequivalent clusters in second order (upper left) and third order (upper right and bottom row). The thickness of the bonds represents the nearest-neighbor correlations $\vec{S}_i \cdot \vec{S}_j$ in the ground state. Summing these contributions yields the ground-state energy per cluster.}
    \label{fig:real space_order23}
\end{figure}

\begin{figure*}[t]
    \centering
    \includegraphics[width=0.19\linewidth]{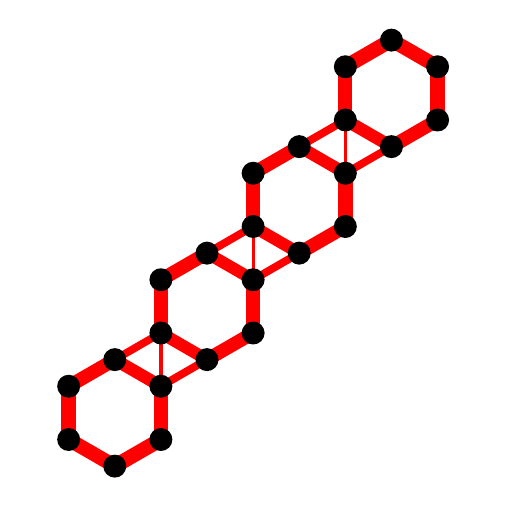}
    \hfill
    \includegraphics[width=0.19\linewidth]{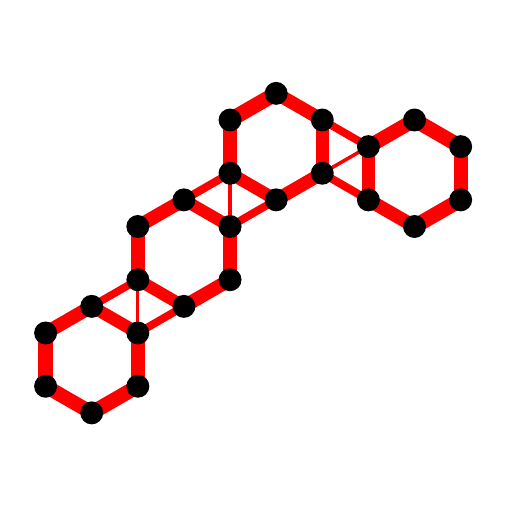}
    \hfill
    \includegraphics[width=0.19\linewidth]{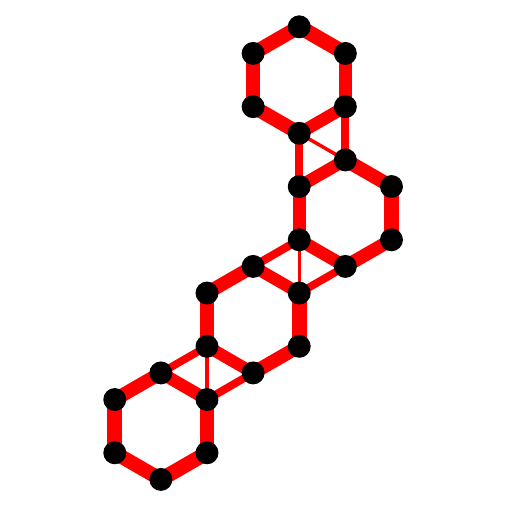}
    \hfill
    \includegraphics[width=0.19\linewidth]{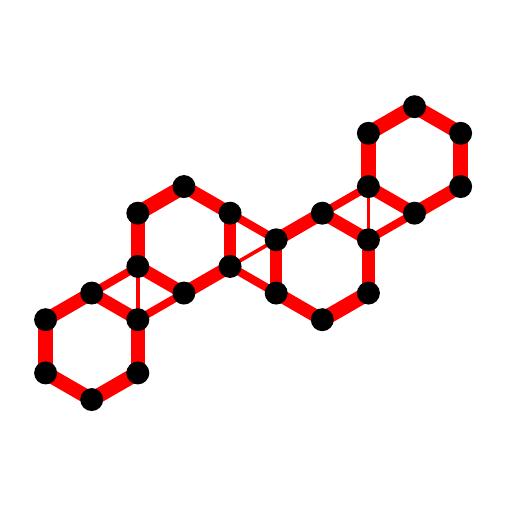}
    \hfill
    \includegraphics[width=0.19\linewidth]{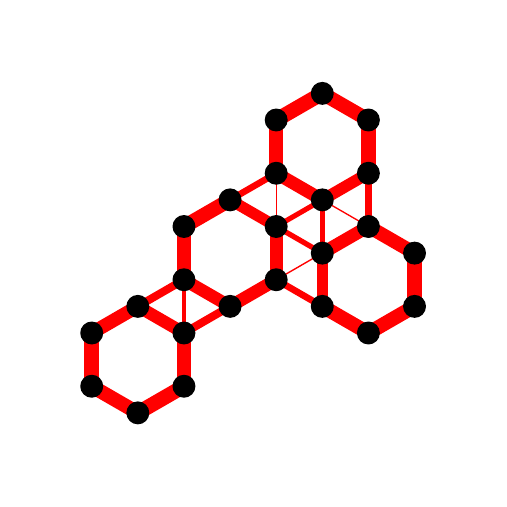}
    \hfill
    \includegraphics[width=0.19\linewidth]{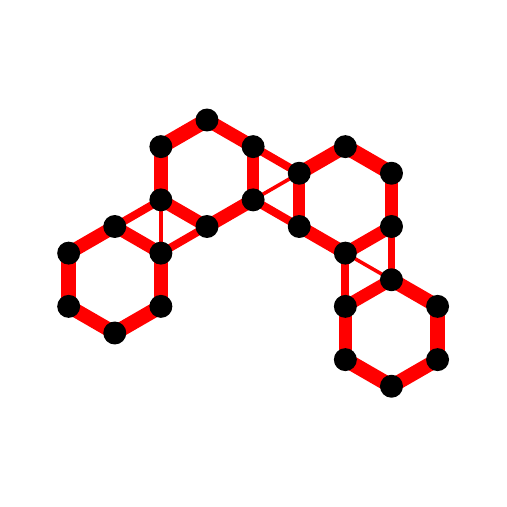}
    \hfill
    \includegraphics[width=0.19\linewidth]{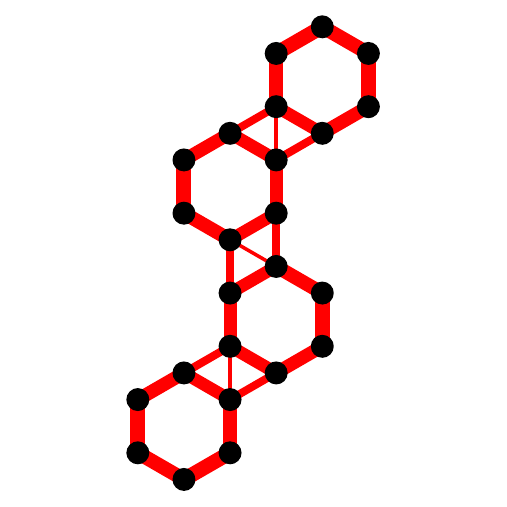}
    \hfill
    \includegraphics[width=0.19\linewidth]{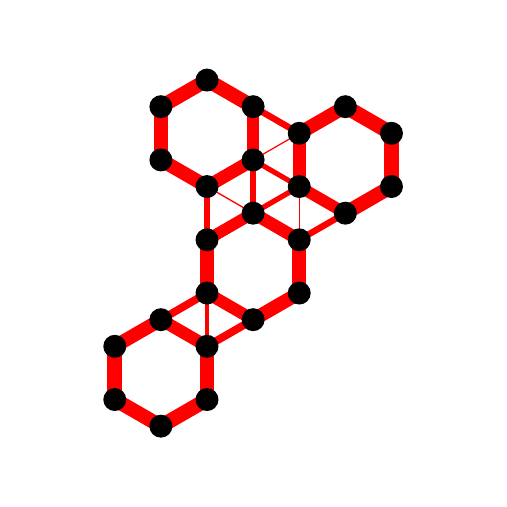}
    \hfill
    \includegraphics[width=0.19\linewidth]{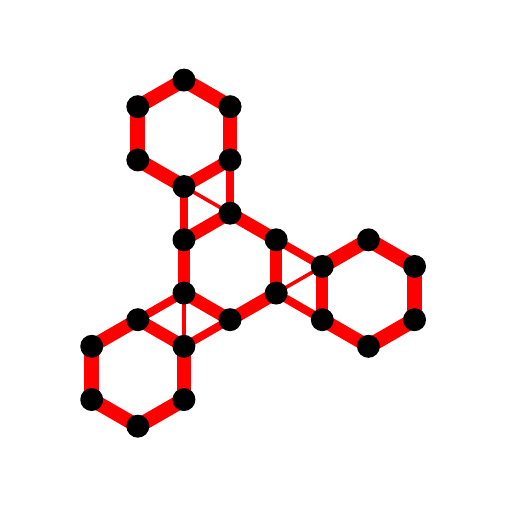}
    \hfill
    \includegraphics[width=0.19\linewidth]{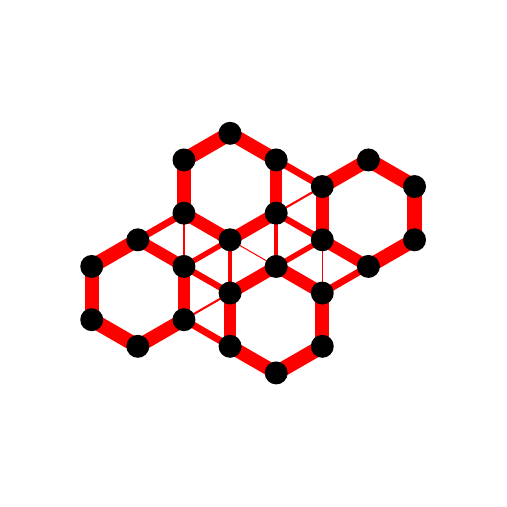}
    \caption{Topologically inequivalent clusters in fourth order. The thickness of the bonds represents the nearest-neighbor correlations $\vec{S}_i \cdot \vec{S}_j$ in the ground state. Summing these contributions yields the ground-state energy per cluster.}
    \label{fig:real space_order4}
\end{figure*}
The convergence of the bare NLCE at zero temperature for the hexagon expansion can be seen by examining the nearest-neighbor correlations in the ground states of clusters at different orders, as shown in \autoref{fig:real space_order23} and \autoref{fig:real space_order4} for orders two, three, and four. These nearest-neighbor correlations directly encode the energy of each cluster. A consistent pattern emerges as the correlations on the three inequivalent bond types (hexagon, triangle, and diagonal) remain robust across all clusters. Hexagon bonds exhibit the strongest correlations, followed by triangle bonds, and the diagonal bonds are the weakest. The expectation values for these three bond types are summarized in \figref{fig:real space_nn}{a} in the main text.

\subsection*{Other expansion}
\begin{figure}[t]
    \includegraphics[width=\columnwidth]{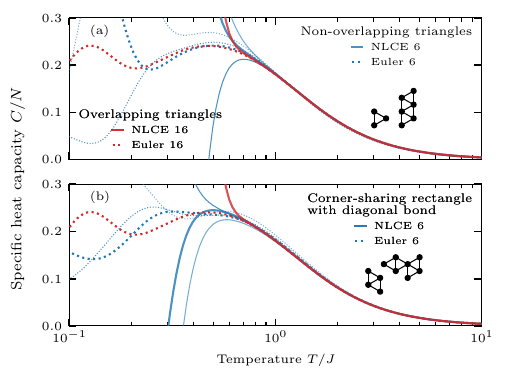}
    \caption{Specific heat capacity obtained from expansions using non-overlapping triangles (a) and corner-sharing rectangles with diagonal-bond expansion (b). The first- and second-order clusters are shown in the inset. The red curve serves as a benchmark, representing the 16th-order overlapping-triangle expansion used in the main text. Dotted lines correspond to the Euler-resummed data, while solid lines show the bare NLCE results. Lower-order results are indicated by thinner, faded lines.}
    \label{fig:NLCE_CV}
\end{figure}

We present additional data for two further expansions, showing the specific heat capacity in \autoref{fig:NLCE_CV}. Panel (a) displays the non-overlapping triangular expansion, with two triangles per unit cell. These triangles are illustrated with red bonds in \autoref{fig:maple_leaf} of the main text. The two triangles are connected by three bonds: two hexagonal bonds and one diagonal bond. Panel (b) shows corner-sharing rectangles with a diagonal bond, with three rectangles per unit cell. The insets display the first- and second-order clusters, respectively. The red curve corresponds to the 16th-order expansion with overlapping triangles used in the main text, shown here for comparison.

While the bare NLCE data (solid lines) exhibits a divergence around $T \approx 0.5\,J$, the Euler resummation (dotted lines) suppresses this divergence, similar to the overlapping-triangle results shown in the main text. We observe significant differences between consecutive orders; nevertheless, the overall trend appears consistent with the data presented in the main text. Both expansions agree well at high temperatures.

\subsection*{Edge-sharing expansion units}
Using edge-sharing expansion units is more advanced because it requires the explicit evaluation of the ``zeroth’’-order contribution. This contribution also becomes necessary for site-sharing expansions as well as non-overlapping expansion units which are subject to local fields.

We begin by discussing the “zeroth’’ order of the site-sharing expansion. Prominent examples include the triangular and tetrahedral expansions on the kagome and pyrochlore lattices, respectively. The zeroth order is the contribution of a single site and we compute the expectation value of the observable of interest on each site. For each cluster generated in the expansion, we must subtract the corresponding single-site expectation value for every site contained in the cluster in order to obtain the correct weight. The total sum over all orders in \autoref{eq:NLCE} must likewise include the single-site contribution. Note that these expectation values may be site-dependent (for example, when different magnetic fields act on the three inequivalent sites of the kagome lattice). Including the zeroth order is only necessary when the single-site expectation value is non-zero. This occurs, for instance, for observables such as the entropy or in the presence of local fields.

For edge-sharing expansion units, the zeroth order includes the contribution of each edge. In the maple-leaf lattice, we work with edge-sharing triangles, of which there are eight inequivalent triangles comprising a total of fifteen inequivalent edges. We evaluate the observable of interest on each of these fifteen edges and subtract the corresponding edge contribution from every generated cluster that includes this edge. The zeroth-order edge contribution is then added to the full sum over all orders in \autoref{eq:NLCE}. Each observable considered here has a non-trivial contribution since the Hamiltonian acts on the edges. Handling the edge contribution carefully becomes especially important when we compute the structure factor, where we need to incorporate the orientation of each edge.
\begin{figure}[t]
    \includegraphics[width=\columnwidth]{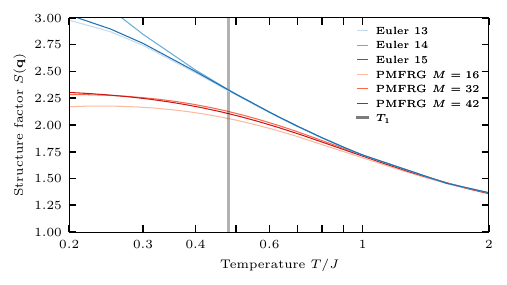}
    \caption{Peak intensity at the $K$ point of the equal-time structure factor computed with PFMRG for increasing numbers of included Matsubara frequencies $M$. The results are compared to the Euler-resummed data from the triangular expansion. The temperature of the first peak, $T_1$, is indicated, corresponding to the third column in \autoref{fig:Sq_highT}.}
    \label{fig:PMFRGscaling}
\end{figure}

\section{Pseudo-Majorana Functional Renormalization Group}\label{app:PMFRG}
To obtain the equal-time structure factor, we sum over all frequencies used in the calculation, as described in \autoref{eq:equal_time_PMFRG} in the main text. In \autoref{fig:PMFRGscaling}, we compare the convergence for different numbers of frequencies $M$ as a function of temperature. The PMFRG data are compared to the NLCE calculation.

While the PMFRG results for $M = 32$ and $M = 42$ are largely consistent with each other, they begin to deviate from the NLCE data for temperatures $T \lesssim 0.9\,J$. Since the state exhibits only short-range correlations, we do not believe that the discrepancy arises from the site cutoff used in our calculations. We attribute the differences to the neglect of higher-order vertex functions within our PMFRG scheme.

We note that we have added the onsite contribution of the equal-time structure factor (which is a simple constant) by hand.

\begin{figure}[t]
    \includegraphics[width=\columnwidth]{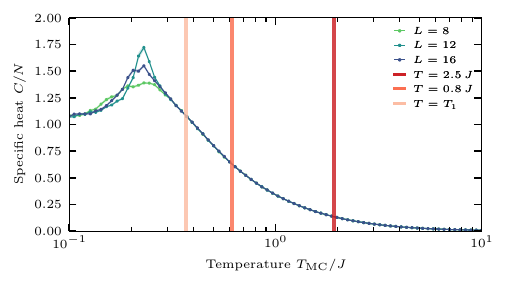}
    \caption{Specific heat capacity obtained from a classical Monte Carlo simulation for different linear system sizes $L$. The number of sites is $N=6L^2$. The corresponding temperatures at which the structure factor is shown in \figref{fig:Sq_highT}{m,n,o} of the main text are marked in red. }
    \label{fig:MC_specific_heat}
\end{figure}
\section{Classical Monte Carlo}\label{app:MC}

We used the \texttt{SpinMC.jl} package~\cite{fbuessen_spin_mc} to perform classical Monte Carlo simulations of $O(3)$ Heisenberg spins for several system sizes. From these simulations, we extracted the specific heat and the equal-time structure factor. The classical model exhibits a non-collinear (canted) planar state without an extensive degeneracy, as discussed in Refs.~\cite{schulenburg:2000,schmalfuss:2002}.

The specific heat capacity shows a single peak located at $\TMC \approx 0.23\,J$ before approaching its low-temperature value, as shown in \autoref{fig:MC_specific_heat}. We evaluate system sizes $L=8,12,16$, corresponding to $384,864,1536$ sites, respectively.

We used the specific-heat data to relate the temperature of the classical ($\TMC$) and quantum ($T$) calculations. For this purpose, we introduced the dimensionless coefficient $\alpha$,
\begin{align}
    T=\alpha \TMC
\end{align}
chosen such that the high-temperature tail of the classical specific heat matches the quantum result. We find $\alpha = 0.77$.

We further rescale the overall intensity of the structure factor of the classical simulation to match the quantum calculation. As a reference point, we matched the high-temperature data from the classical simulation at $T=2.5\,J$ to the high-temperature result from the triangular expansion. This scaling yields an excellent approximation at lower temperatures, $T=0.8\,J$ and $T=T_1$, which rely only on the high-temperature rescaling.

\begin{figure}[t]
    \includegraphics[width=\columnwidth]{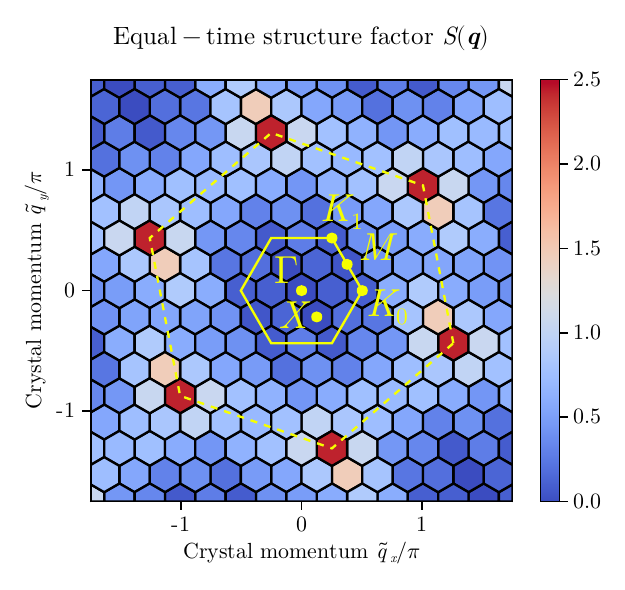}
    \caption{\label{fig:appendix:ED-structureFactor}Spin structure factor in the ground state of the $N=36$ cluster based on ED. Reciprocal vectors resolved by the cluster are drawn as colored hexagons without any interpolation.
    The solid yellow line highlights the first Brillouin zone of the maple-leaf lattice and momenta resolved by the cluster are drawn. The Brillouin zone of the underlying triangular lattice is shown as the dashed line.
    Contrary to the previous figures, we here employed a rotated lattice convention, which is why we label the momentum variable with $\tilde{q}$ instead of $q$.
    }
\end{figure}
\section{Zero-temperature structure factor ED}\label{app:ED-structure-Factor}
As a cross-check to our results from \autoref{sec:StructureFactor}, we present the structure factor obtained from ED on the most symmetric 36-site cluster in \autoref{fig:appendix:ED-structureFactor}. The used cluster can be seen in Refs.~\cite{farnell:2011, schmalfuss:2002}.

This cluster resolves $\Gamma$, $K$, $M$ and even $X$ points in reciprocal space, but not all their associated little co-groups.
This shortcoming is reflected in the structure factor having a mere $C_2$ instead of the $C_6$ symmetry expected from the maple-leaf lattice.
Nevertheless, our ED results are in excellent agreement with those discussed in \autoref{sec:StructureFactor}, as well as with structure factors obtained from other numerical methods~\cite{gresista:2023, beck:2024, schmoll_bathing_2025, ghosh:2025}. The data clearly show peaks at the $K$ points of the underlying triangular lattice, along with a subtle twist around the peaks. The intensity matches qualitatively.

\end{document}